\documentstyle[12pt,epsf]{ioplppt}
\begin{document}
\newcommand{\be}{\begin{equation}\label}
\newcommand{\ee}{\end{equation}}
\newcommand{\bea}{\begin{eqnarray}\label}
\newcommand{\eea}{\end{eqnarray}}
\newcommand{\ul}{\underline}
\newcommand{\vmu}{\ul{\mu}}
\newcommand{\vPhi}{\ul{\Phi}}
\newcommand{\vPsi}{\ul{\Psi}}
\newcommand{\mL}{\ul{\ul{L}}}
\newcommand{\Id}{\ul{\ul{1}}}
\newcommand{\vu}{\ul{u}}
\newcommand{\vv}{\ul{v}}
\newcommand{\vw}{\ul{w}}
\newcommand{\vx}{\ul{x}}
\newcommand{\vy}{\ul{y}}
\newcommand{\vC}{\ul{C}}
\newcommand{\vD}{\ul{D}}
\newcommand{\vE}{\ul{E}}
\newcommand{\vF}{\ul{F}}
\newcommand{\eps}{\varepsilon}
\newcommand{\vG}{\ul{\Gamma}}
\newcommand{\vg}{\ul{\gamma}}
\newcommand{\vCt}{\tilde{\vC}}
\newcommand{\vDt}{\tilde{\vD}}
\newcommand{\db}{\delta \beta}
\newcommand{\Det}{\mbox{Det}}
\renewcommand{\Re}{\mbox{Re}}
\renewcommand{\Im}{\mbox{Im}}
\jl{1}
\title{On the degenerated soft--mode 
instability}[Degenerated soft--mode instability]
\author{
Wolfram Just\dag, 
Frank Matth\"aus\ddag, 
Herwig Sauermann\ddag}
\address{\dag\ Max--Planck Institute for Physics of Complex Systems,
N\"othnizer Stra\ss e 38, D--01187 Dresden, Germany}
\address{\ddag Technical University Darmstadt, Hochschulstra\ss e 8,
D--64289 Darmstadt, Germany}
\begin{abstract}
We consider instabilities of a single mode with finite wavenumber
in inversion symmetric {  spatially one dimensional}
systems, where
the character of the bifurcation changes from sub-- to supercritical
behaviour. Starting from a general equation of motion the full amplitude
equation is derived systematically and
formulas for the dependence of the coefficients
on the system parameters are obtained. 
We emphasise the importance of nonlinear derivative
terms in the amplitude equation for the behaviour in the vicinity of the
bifurcation point. Especially the numerical values of the corresponding 
coefficients determine the region of coexistence between the stable
trivial solution and stable spatially periodic patterns. Our approach
clearly shows that similar considerations fail for the case
of oscillatory instabilities.
\end{abstract}
\maketitle
\section{Introduction}
Pattern formation in systems of large size has attracted 
much research interest in recent years. Especially the fact that
many aspects, at least of one--dimensional systems or of quasi 
one--dimensional patterns, can be described by reduced equations of motion
has allowed for linking quite different fields of physics 
(cf.~\cite{CrHo} and references therein). To some extent the approach 
strongly parallels the normal
form calculations in low dimensional dynamical systems \cite{GuHo}, 
The reduced equations for simple instabilities, i.~e.~the 
Ginzburg--Landau equation is well established and its derivation 
can even be found in textbooks \cite{Mann}. However, at least from the 
general point of view, less is known if additional constraints are 
imposed on the instability, that means if
higher order codimension bifurcations are considered.

We are here concerned with instabilities {  in spatially one dimensional
systems} where a single mode 
becomes unstable with respect to a wavenumber $q_c$ due to an
eigenvalue zero in the spectrum. Such a situation occurs
generically in inversion symmetric situations and we henceforth 
consider such systems. Let 
$\lambda (k,\vmu)$ denote the corresponding critical
eigenvalue in dependence on the wavenumber $k$
and the system parameters $\vmu$. It obeys
\be{s1-a}
\lambda (q_c,\vmu)=0,\quad
\left. \partial_k \lambda (k,\vmu) 
\right|_{k=q_c}=0, \quad
\left. \partial^2_k \lambda (k,\vmu) \quad
\right|_{k=q_c}<0 \quad .
\ee
These conditions trace a codimension one set, i.~e.~a hypersurface
in the parameter space on which the instability occurs. It is well
established, even from a rigorous point of view \cite{Coll}, that the dynamics
near such an instability is governed by a slowly varying envelope which obeys
a real Ginzburg--Landau equation. Whether the instability is 
sub-- or supercritical, i.~e.~whether the amplitude saturates in the vicinity 
of the instability, depends on the sign of the cubic term. The transition from
sub-- to supercriticality, i.~e.~the change of the sign, leads to a codimension
two bifurcation\footnote{  Sometimes such a bifurcation point is called
a tricritical point, in order to distinguish it from a codimension two
bifurcation caused by the degeneracy of two distinct modes.}. 
It is of course
contained in the bifurcation set determined by 
eqs.(\ref{s1-a}). Such instabilities, which for example are relevant in
the hydrodynamic context (cf.~\cite{BeSw} for a recent reference) 
are at the centre of interest of our contribution.

From pure symmetry considerations 
the structure of the reduced amplitude equation may be 
fixed, taking the translation and inversion 
symmetry into account
\be{s1-b}
\partial_{\tau} \bar{A} = \bar{\eta} \bar{A} + D 
\partial^2_{\xi} \bar{A} + \bar{r} |\bar{A}|^2 \bar{A} 
+ s|\bar{A}|^4 \bar{A} + \i g |\bar{A}|^2 \partial_{\xi} \bar{A} 
+ \i d \bar{A}^2 \partial_{\xi} \bar{A}^* \quad .
\ee
However, these considerations do not tell us whether such an equation
is valid at all, and how the coefficients depend on the actual parameters
of the underlying equations of motion. We present the
complete derivation of the amplitude equation (\ref{s1-b}) 
starting from a general equation of motion, even if the method
is in principle well established in the hydrodynamic context.
But our approach is purely algebraic
and has the advantage, that the results can be applied immediately
to quite different physical situations. As a by--product we remark
that for the similar hard--mode case a comparable approach fails,
in contrast to statements in the literature. Finally we will dwell on
some properties  of eq.(\ref{s1-b}), since a complete discussion 
is difficult to find in the literature, despite the fact that 
related results from different points of view can be found quite 
frequently \cite{DeBr,SaHo,EcIo,DoEc}.
\section{Derivation of the amplitude equation}\label{sec2}
\subsection{Notation}\label{sec21}
We suppose that the basic equation of motion for the $N$--component
real field $\vPhi(x,t)$ is cast into the form
\be{s2-a}
\partial_t \vPhi= {\cal L}_{\vmu} \vPhi + {\cal N}[\vPhi;\vmu]
\ee
such that the trivial translation invariant stationary state is given by
$\vPhi\equiv 0$ \footnote{  In particular we concentrate on situations where
boundary conditions play no significant role, so that we can consider formally
systems of infinite extent.}. 
For the linear operator, which determines the instability
of this state, we allow for an expression as general as possible, i.~e.
\be{s2-b}
{\cal L}_{\vmu} \vPsi = \sum_{\alpha} \mL_{\alpha}(\vmu) 
\partial^{\alpha}_x \vPsi \quad .
\ee
Using plane wave solutions $\vu \exp(\i k x)$ the eigenvalue problem is
completely determined by the $N\times N$ matrices
\be{s2-c}
\mL(k;\vmu) = \sum_{\alpha} \mL_{\alpha}(\vmu) (\i k)^{\alpha}
\ee
according to
\be{s2-ca}
\mL(k;\vmu) \vu_k(\vmu)= \lambda (k,\vmu)
\vu_k(\vmu) \quad .
\ee
We denote by
$\lambda$ the eigenvalue branch with maximal real part, which
obeys eqs.(\ref{s1-a}) at the instability. In our case the inversion symmetry
guarantees that all quantities are real valued.

For the nonlinear part we employ an expansion in powers of the field
according to
\be{s2-d}
{\cal N}(\vPsi;\vmu) = {\cal N}_2(\vPsi;\vmu)+{\cal N}_3(\vPsi;\vmu)+
{\cal N}_4(\vPsi;\vmu)+ {\cal N}_5(\vPsi;\vmu)+\cdots
\ee
where
\bea{s2-e}
{\cal N}_2(\vPsi;\vmu) = \sum_{\alpha,\beta} \vC^{(\alpha \beta)}_{\vmu}
\{\partial^{\alpha}_x \vPsi,\partial^{\beta}_x \vPsi\}\label{s2-ea}\\
{\cal N}_3(\vPsi;\vmu) = \sum_{\alpha,\beta,\gamma} 
\vD^{(\alpha \beta \gamma)}_{\vmu}
\{\partial^{\alpha}_x \vPsi,\partial^{\beta}_x \vPsi,
\partial^{\gamma}_x \vPsi\}\label{s2-eb}
\eea
denote the most general expressions of second and third order, 
with vector valued bi-- and trilinearforms $\vC^{(\alpha \beta)}$ and
$\vD^{(\alpha \beta \gamma)}$. Written in components they read for example 
\be{s2-f}
\left( \vC^{(\alpha \beta)}_{\vmu } \{\vu,\vv \} \right)_m
= \sum_{k,l} c_{\vmu; m k l}^{(\alpha \beta)} u_k v_l \quad .
\ee
The contributions of order four and five are understood in the same way
using the notation $\vE^{(\alpha \beta \gamma \delta)}$ and
$\vF^{(\alpha \beta \gamma \delta \epsilon)}$ for the corresponding
multilinearforms\footnote{In addition the symmetry properties
$\vC^{(\alpha \beta)}\{\vu,\vv\}=\vC^{(\beta \alpha)}\{\vv,\vu\}$,
$\vD^{(\alpha \beta \gamma)}\{\vu,\vv,\vw\}=
\vD^{(\gamma \alpha \beta )}\{\vw,\vu,\vv,\}=\ldots$, $\dots$ are employed
in what follows.}. 
In the subsequent analysis it is necessary to evaluate
the nonlinearities if plane waves are inserted for the field. To be specific
only waves with the multiples of the critical wavenumber $q_c$ will occur.
In such a case all the nonlinearities are expressed in terms of the 
abbreviations (cf.~eq.(\ref{s2-c}))
\bea{s2-g}
\fl
\vC_{m n}(\vu, \vv ; \vmu) := \sum_{\alpha \beta} (\i m q_c)^{\alpha}
(\i n q_c)^{\beta} \vC^{(\alpha \beta)}_{\vmu}
\{\vu,\vv\} \label{s2-ga}\\
\fl
\vD_{l m n}(\vu, \vv,\vw ; \vmu) 
:= \sum_{\alpha \beta \gamma} (\i l q_c)^{\alpha}
(\i m q_c)^{\beta}  (\i n q_c)^{\gamma} 
\vD^{(\alpha \beta \gamma)}_{\vmu}
\{\vu,\vv,\vw\} \label{s2-gb}\\
\fl
\vE_{k l m n}(\vu, \vv,\vw ,\vx; \vmu) 
:= \sum_{\alpha \beta \gamma \delta} (\i k q_c)^{\alpha}
(\i l q_c)^{\beta}  (\i m q_c)^{\gamma} (\i n q_c)^{\delta} 
\vE^{(\alpha \beta \gamma \delta)}_{\vmu}
\{\vu,\vv,\vw,\vx\} \label{s2-gc}\\
\fl
\vF_{j k l m n}(\vu, \vv,\vw, \vx, \vy; \vmu) 
:= \sum_{\alpha \beta \gamma \delta \epsilon } (\i j q_c)^{\alpha}
(\i k q_c)^{\beta}  (\i l q_c)^{\gamma}  (\i m q_c)^{\delta} 
(\i n q_c)^{\epsilon} 
\vF^{(\alpha \beta \gamma \delta \epsilon)}_{\vmu}
\{\vu,\vv,\vw,\vx,\vy\} 
\label{s2-gd}
\eea
which are
frequently used in what follows.
\subsection{Weakly nonlinear analysis}
Suppose that at $\vmu=\vmu_c$ a degenerated soft--mode instability
occurs. Let $\vu_c \exp(\i q_c x)$ denote the marginally stable mode,
i.~e.~$\vu_c=\vu_{q_c}(\vmu_c)$ is the nulleigenvector 
of the matrix (\ref{s2-c}) at
$\vmu=\vmu_c$ and $k=q_c$. In the vicinity of this parameter value, i.~e.~for
\be{s2-h}
\vmu=\vmu_c + \eps \vmu^{(1)} + \eps^2 \vmu^{(2)} + \cdots
\ee
the solution of the full equation is expanded as
\bea{s2-i}
\vPhi(x,t)=\eps^{1/2} \vPhi_1 +\eps \vPhi_2+ \eps^{3/2} \vPhi_3+
\eps^2 \vPhi_4+\eps^{5/2} \vPhi_5 + \cdots \label{s2-ia}\\
\vPhi_1= \vu_c \exp(\i q_c x) A(\tau_1,\tau_2,\ldots,\xi_1,\xi_2,\ldots)+ c.c
\label{s2-ib}
\eea
where $\eps$ denotes a dimensionless smallness parameter. 
As usual the dynamics of the complex valued
amplitude $A$, which possesses a slowly varying
space--time dependence on the scales $\tau_k=\eps^k t$, $\xi_k=\eps^k x$,
will be determined by the secular conditions of the expansion.
However, before we proceed let us comment on the choice of the expansion. 
The actual expansion parameter is given by $\eps^{1/2}$ and
for completeness the slow scales $\eps^{1/2} t$ and $\eps^{1/2} x$ should 
also be
taken into account. But as will become obvious from the following 
considerations these scales drop and do not contribute. The same conclusion
holds for the expression (\ref{s2-h}), where $\vmu^{(1,2)}$ act as unfolding
parameters. In addition, since $\vmu^{(1)}$ will
be confined along the codimension one set of soft--mode bifurcations, the
real unfolding of the bifurcation occurs at the order $\Or (\eps^2)$.
Hence the scaling of the amplitude in eq.(\ref{s2-ia}) coincides
with the scaling in the corresponding
spatially homogeneous situation. Nevertheless, we stress
again that the inclusion of all terms of orders $\Or(\eps^{k/2})$ yields
the same results as presented below.

If one inserts the expansion (\ref{s2-h}) into the definitions
(\ref{s2-b}) and (\ref{s2-d}) one obtains
\be{s2-j}
{\cal L}_{\vmu}= {\cal L}+\eps {\cal L}^{(1)} +\eps^2 {\cal L}^{(2)} +\ldots
\ee
and
\be{s2-k}
{\cal N}_k(\vPsi;\vmu) ={\cal N}_k(\vPsi) +\eps {\cal N}^{(1)}_k(\vPsi) +
\dots \quad .
\ee
Here the terms of higher order contain the parameters $\vmu^{(1,2)}$.
In order to simplify the notation we do not introduce a different symbol
for the contributions of order zero, but just skip the argument $\vmu$.
Even this convention introduces a slight abuse of notation, it is
henceforth understood that the corresponding expressions are evaluated at
$\vmu=\vmu_c$, e.~g.~${\cal L}:={\cal L}_{\vmu_c}$. Analogous expansions hold 
for quantities like eq.(\ref{s2-c}) or (\ref{s2-g})--(\ref{s2-gd}). 
In particular the matrix $\mL(k)=\mL(k;\vmu_c)$ determines the critical 
mode and $\mL^{(1)}(k)=(\vmu^{(1)}\partial_{\vmu})\mL(k;\vmu)|_{\vmu=\vmu_c}$.

The following steps are now like for the usual codimension one case
and can in principle be found in
textbooks. If one inserts the expansion (\ref{s2-i}) into the equation
of motion (\ref{s2-a}), takes eqs.(\ref{s2-j}) and (\ref{s2-k})
into account and performs the derivatives with respect to all scales, then
one obtains at each order in $\eps^{k/2}$ an inhomogeneous
linear equation determining $\vPhi_k$
\be{s2-l}
\partial_t \vPhi_k = {\cal L} \vPhi_k + 
\sum_{n} \exp(\i n q_c x) \vw_n  \quad .
\ee
The inhomogeneous part typically contains Fourier modes with
integer multiples of the critical mode, and
the slow scales are just considered as fixed parameters.
If $\vv_c$ denotes the 
left nulleigenvector of the critical mode, i.~e.~$\vv_c^* \mL(q_c)=0$, then
the condition that the solution of eq.(\ref{s2-l}) does not become
secular reads
\be{s2-m}
\langle \vv_c | \vw_{n=1}\rangle=0 \quad .
\ee
Here the brackets denote the usual scalar product. Now
the solution, discarding exponentially decaying transients, reads
\bea{s2-n}
\vPhi_k &=& -\sum_{n\neq \pm 1} \mL(n q_c)^{-1} \vw_n 
\exp(\i n q_c x)\nonumber \\
&-& \left( \mL_c^{-1} \vw_{n=1} \exp(\i q_c x) 
+ \vu_c Z(\tau_1,\ldots,\xi_1,\ldots) \exp(\i q_c x) + c.c. \right)
\eea
where the coefficient $Z$ of the solution of the homogeneous equation
may depend on the slower scales. $\mL_c^{-1}$ denotes the
inverse on the subspace omitting the critical mode, i.~e.
\be{s2-o}
\mL_c^{-1} \mL(q_c) = \mL(q_c) \mL_c^{-1}  
=\Id - |\vu_c\rangle \langle \vv_c|,\quad
\mL_c^{-1} \vu_c= 0, \quad \vv_c^* \mL_c^{-1}=0 \quad .
\ee
The reason that we reiterate this scheme here is twofold. On the one hand we
would like to give the reader the chance, within the amount of
formalism to relax 
with a passage, which he or she of course knows quite well. On the other 
hand the textbooks mentioned above usually stop with such general 
considerations or specialise to certain conditions which may not be shared
by the model under consideration. Here we will continue with the most general
equation of motion, even if we have to proceed to the fifth order. We
demonstrate that the explicit evaluation is not at all horrible
within a suitable notation.

At the order $\Or(\eps^{1/2})$ eq.(\ref{s2-l}) just yields the eigenvalue
equation for the critical mode (cf.~eq.(\ref{s2-ca})). 
At order $\Or(\eps)$ the quadratic nonlinearity
contributes nonresonant Fourier modes $\pm 2 q_c,0$ to the inhomogeneous part
(cf.~eq.(\ref{a-a})), so that no nontrivial secular condition occurs. 
The solution (\ref{s2-n}) reads in this case
\be{s2-p}
\vPhi_2= 2 \vG_{20} |A|^2 
+\left( \vG_{22} \exp(\i 2 q_c x) A^2 + \vu_c \exp(\i q_c x) B + c.c.\right)
\ee
where the amplitude $B$ of the homogeneous solution depends on the slower 
scales and the abbreviations (\ref{a-b}) are introduced. 

\begin{figure}
\begin{center}
\mbox{\epsffile{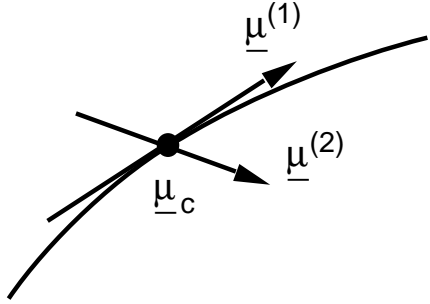}}
\end{center}
\caption[ ]{Diagrammatic view of the parameter space in the vicinity
of the degenerated soft--mode bifurcation point $\vmu_c$. The solid
line indicates the soft--mode bifurcation manifold and $\vmu^{(1,2)}$
the unfolding vectors (cf.~eq.(\ref{s2-h})). \label{fig-a}}
\end{figure}

At the third order $\Or(\eps^{3/2})$ the quadratic 
and cubic nonlinearity contribute as well as the linear operators ${\cal L}$
and $\partial_t$, if the derivatives with respect to the slow scales are 
performed. The complete inhomogeneous part of eq.(\ref{s2-n}) is 
for convenience given in eq.(\ref{a-c}). The secular condition (\ref{s2-m})
yields
\be{s2-q}
0 = - \langle \vv_c | \vu_c \rangle  \partial_{\tau_1} A  + 
\langle \vv_c | \mL^{(1)}(q_c)\ \vu_c \rangle A
+ \rho_3 A |A|^2
\ee
if we equate the coefficient of $\partial_{\xi_1} A$ to zero with
the help of eq.(\ref{s1-a}). In the usual codimension one case we would now
end up with the Ginzburg--Landau equation. Here however we require that the
coefficient of the cubic term vanishes\footnote{Indices with an overbar 
denote negative values $\bar{n}:=-n$.} 
\be{s2-r}
\fl \rho_3(q_c;\vmu_c):=
\langle \vv_c | 2\vC_{2\bar{1}} (\vG_{22}, \vu_c) +
4\vC_{10} (\vu_c, \vG_{20}) + 3\vD_{11\bar{1}} (\vu_c, \vu_c, \vu_c)  
\rangle \stackrel{!}{=} 0 \quad .
\ee
Together with the condition (\ref{s1-a}) this equation determines
the codimension two bifurcation manifold. Since the secular condition 
(\ref{s2-q}) has to yield a finite and nonvanishing solution, we finally 
have to require that both of the two remaining terms vanish separately. 
Hence we are left with
\be{s2-s}
\partial_{\tau_1} A= 0
\ee
and
\be{s2-sa}
\vmu^{(1)} \partial_{\vmu} \lambda (q_c,\vmu)|_{\vmu=\vmu_c}=0
\ee
if we take into account, that the matrix element in eq.(\ref{s2-q}) can
be expressed in terms of a directional derivative owing to the
definitions (\ref{s2-h}) and (\ref{s2-j}).
The condition (\ref{s2-sa}) has a simple geometrical interpretation 
(cf.~fig.\ref{fig-a}).
If we take the total derivative of eqs.(\ref{s1-a}) with respect to $\vmu$
along a direction in the codimension one bifurcation manifold, 
i.~e.~we take the dependence of the critical wavenumber on $\vmu$ into 
account, we are exactly left with eq.(\ref{s2-sa}). Hence the secular 
condition fixes the parameter variation $\vmu^{(1)}$ at order $\Or(\eps)$ 
in such a way that only variations within the soft--mode instability
manifold are permitted. The full parameter unfolding is obtained
at higher order. By the way we remark that similar considerations
exclude the spatial scale $\eps^{1/2} x$ from the perturbation expansion.
For the solution we now get the result
\bea{s2-t}
\fl
\vPhi_3 &=&  \left[
\vG_{31} A |A|^2 - \i \vg_{31}^a \partial_{\xi_1}A +
\vg_{31}^b A \right] \exp(\i q_c x) + \vG_{33} \exp(\i 3 q_c x) A^3 
\nonumber\\
\fl
&+& 2 \vG_{20} A B^* + 2\vG_{22}\ \exp(\i 2 q_c x) A B  +
\vu_c \exp(\i q_c x) C + c.c.
\eea
using the abbreviations (\ref{a-da})--(\ref{a-dd}).

At the order $\Or(\eps^2)$ also the parameter dependence of the
nonlinearities (cf.~eq.(\ref{s2-k})) contributes. The inhomogeneous part,
which is given in eq.(\ref{a-e}) up to Fourier modes $2 q_c$
yields for the secular condition (\ref{s2-m}), taking the secular 
condition (\ref{s2-sa}) of the preceding order into account
\be{s2-u}
0 = -\langle \vv_c | \vu_c \rangle  \partial_{\tau_1} B
+ \rho_3 \left[2 |A|^2 B  +  A^2 B^* \right]  \quad .
\ee
By virtue of the higher order codimension condition (\ref{s2-r}) we
are left with
\be{s2-v}
\partial_{\tau_1} B=0 \quad .
\ee
For the solution at this order we have, if we restrict to Fourier modes
$\pm2q_c,0$ which will become resonant at the next order
\bea{s2-w}
\fl
\vPhi_4 &=& \vG_{40} |A|^4  + 2 \vg^b_{40} |A|^2 + 2 \vG_{20}  |B|^2\ 
+ \left( 2 \i \vg^a_{40}  A \partial_{\xi_1} A^*  +  2 \vG_{20} A C^* \right .
\nonumber \\
\fl
&+&
\left. 
\exp(\i 2 q_c x) \left[ \vG_{42} A^2 |A|^2 + 2 \i \vg^a_{42} A 
\partial_{\xi_1}A
+ 2 \vG_{22} A C  + \vg^b_{42} A^2 +  \vG_{22} B^2 \right] 
+ c.c. \right) +  \ldots \quad .
\eea
The coefficients are given by eqs.(\ref{a-fa})--(\ref{a-ff}).

We now plug in all results to compute the secular condition at the
order $\Or(\eps^{5/2})$ and obtain, taking eqs.(\ref{s1-a}), (\ref{s2-s}),
(\ref{s2-sa}), and (\ref{s2-v}) into account
\bea{s2-x}
\fl
\langle \vv_c | \vu_c \rangle \left[
- \partial_{\tau_2} A 
+ \eta A  + D \partial_{\xi_1}^2 A + \i c \partial_{\xi_1} A
+ r A |A|^2 + s A |A|^4 + \i g |A|^2 \partial_{\xi_1}A 
+ \i d A^2\partial_{\xi_1} A^* \right] \nonumber \\
\fl
+ \rho_3 \left[ 2 |A|^2 C + A^2 C^* +  A^* B^2 + 2 A |B|^2 \right]
- \langle \vv_c | \vu_c \rangle \partial_{\tau_1} C = 0 \quad .
\eea
Thanks to the higher order codimension condition (\ref{s2-r}) the nonlinear
terms which couple the different amplitudes vanish. 
Furthermore the last summand,
being solely dependent on the scale $\tau_1$ has to vanish too, in order to
avoid a secular contribution. If we introduce $\bar{A}:=
\exp[\i c \xi_1/(2D)] A$ to eliminate the linear derivative term, we
are left with the closed amplitude equation (\ref{s1-b}), where
\be{s2-xa}
\bar{\eta}:=\eta+c^2/(4D) , \quad \bar{r}:=r+c(g-d)/(2D) \quad .
\ee
It is worth to mention that our formalised approach has enabled us
to incorporate the higher order codimension condition 
at all steps in the perturbation expansion.
\subsection{Coefficients}
In addition we have obtained the general
microscopic expressions for the coefficients.
We use the notation introduced in section \ref{sec21} and the abbreviations of
the appendix.

The linear unfolding parameter reads
\bea{s2-y}
\langle \vv_c | \vu_c \rangle \eta := \langle \vv_c | \mL^{(1)}(q_c)
\vg_{31}^b \rangle + \langle \vv_c | \mL^{(2)}(q_c)\vu_c \rangle 
\nonumber \\
= \langle \vv_c | \vu_c \rangle \left[ 1/2  (\vmu^{(1)} \partial_{\vmu})^2 
\lambda (q_c,\vmu)|_{\vmu=\vmu_c} +
(\vmu^{(2)} \partial_{\vmu}) \lambda (q_c,\vmu)|_{\vmu=\vmu_c} \right]
\quad ,
\eea
where the last expression follows from the definitions (\ref{s2-h})
and (\ref{s2-j}) 
straightforwardly. Hence the linear unfolding contains a contribution from
the curvature of the bifurcation manifold and one from the transversal
intersection. 

The diffusion constant is given by
\be{s2-z}
\fl
\langle \vv_c | \vu_c \rangle D := -1/2 \langle \vv_c | \mL''(q_c) \vu_c 
\rangle - \langle \vv_c | \mL'(q_c) \vg_{31}^a \rangle = -1/2 
\langle \vv_c|\vu_c \rangle \partial_k^2
\lambda (k,\vmu_c)|_{k=q_c} \quad .
\ee

For the linear derivative term we have obtained
\bea{s2-A}
\langle \vv_c | \vu_c \rangle c &:=&
\langle \vu_c | -\mL'(q_c) \vg_{31}^b -\mL^{(1)}(q_c) \vg_{31}^a
-(\mL^{(1)})'(q_c) \vu_c \rangle \nonumber \\
&=& - \langle \vv_c | \vu_c \rangle
\partial_k (\vmu^{(1)} \partial_{\vmu}) \lambda (k,\vmu)
|_{k=q_c,\vmu=\vmu_c} \quad .
\eea
In view of the relations (\ref{s1-a}) the derivative can be expressed also
in terms of the change of the critical wavenumber along the 
soft--mode bifurcation manifold.

The cubic unfolding coefficient reads
\bea{s2-B}
\fl \langle \vv_c | \vu_c \rangle r &:=&
\langle \vv_c | \mL^{(1)}(q_c) \vG_{31} 
+ 2\vC_{2\bar{1}} (\vG_{22}, \vg_{31}^b) 
+2\vC_{2\bar{1}} (\vg_{42}^b, \vu_c) 
\nonumber \\
\fl
&+& 
4 \vC_{10} (\vg_{31}^b, \vG_{20}) 
+ 4\vC_{10} (\vu_c, \vg_{40}^b) 
+ 6 \vD_{11\bar{1}} (\vg_{31}^b, \vu_c, \vu_c) 
+ 3 \vD_{11\bar{1}} (\vu_c, \vu_c, \vg_{31}^b) 
\nonumber \\
\fl
&+& 
2\vC_{2\bar{1}}^{(1)} (\vG_{22}, \vu_c) 
+ 4 \vC_{10}^{(1)} (\vu_c, \vG_{20}) 
+ 3 \vD_{11\bar{1}}^{(1)} (\vu_c, \vu_c, \vu_c) \rangle \quad .
\eea
If we use here the representations (\ref{a-dd}), (\ref{a-ha}), and 
(\ref{a-hb}) as well as the higher order codimension relation 
(\ref{s2-r}) the expression 
simplifies to
\be{s2-C}
\langle \vv_c | \vu_c \rangle r = (\vmu^{(1)}\partial_{\vmu})
\rho_3 (q_c;\vmu)|_{\vmu_c=\vmu}
\ee
if we define the object on the right hand side by eq.(\ref{s2-r}) but
evaluated with the full parameter dependent nonlinearities (\ref{s2-ga}),
(\ref{s2-gb}) and eigenvectors (cf.~eq.(\ref{s2-ca})).

The evaluation of the coefficient of the quintic term yields
\bea{s2-D}
\fl
\langle \vv_c | \vu_c \rangle s &:=& \langle \vv_c | 
2 \vC_{10} (\vu_c, \vG_{40}) 
+ 2 \ul{C}_{2\bar{1}} (\vG_{22}, \vG_{31}) 
+ 2 \vC_{2\bar{1}} (\vG_{42}, \vu_c)
+ 2\vC_{3\bar{2}} (\vG_{33}, \vG_{22}) \nonumber \\
\fl
&+&  4 \vC_{10} (\vG_{31}, \vG_{20}) 
+ 6 \vD_{11\bar{1}} (\vG_{31}, \vu_c, \vu_c) 
+ 3 \vD_{11\bar{1}} (\vu_c, \vu_c, \vG_{31}) \nonumber \\
\fl
&+& 3\vD_{3\bar{1}\bar{1}} (\vG_{33}, \vu_c, \vu_c) 
+ 6 \vD_{21\bar{2}} (\vG_{22}, \vu_c, \vG_{22}) 
+ 12 \vD_{100} (\vu_c, \vG_{20}, \vG_{20}) \nonumber \\
\fl  
&+& 12 \vD_{20\bar{1}} (\vG_{22}, \vG_{20}, \vu_c) 
+ 4 \vE_{111\bar{2}} (\vu_c, \vu_c, \vu_c, \vG_{22}) 
+ 12 \vE_{21\bar{1}\bar{1}} (\vG_{22}, \vu_c, \vu_c, \vu_c) \nonumber \\
\fl
&+& 24 \vE_{110\bar{1}} (\vu_c, \vu_c, \vG_{20}, \vu_c) 
+ 10 \vF_{111\bar{1}\bar{1}}(\vu_c, \vu_c, \vu_c, \vu_c, \vu_c)
\rangle \quad .
\eea
This expression is a genuine term of the fifth order and cannot be
reduced further.

For the normal derivative term the coefficient reads
\bea{s2-E}
\fl
\langle \vv_c | \vu_c \rangle g &:=& \langle \vv_c |
-2 \mL'(q_c)_{(q_c)}  \vG_{31} 
-4 \vC_{10} (\vu_c,  \vg^a_{40}) 
+ 4 \vC_{2\bar{1}} (\vg^a_{42}, \vu_c) \nonumber \\
\fl
&-& 4 \vC_{10} (\vg^a_{31}, \vG_{20}) - 4 \vCt_{01} (\vG_{20}, \vu_c) 
- 4 \vCt_{10} (\vu_c, \vG_{20}) - 4 \vCt_{2\bar{1}} (\vG_{22}, \vu_c) 
\nonumber \\
\fl
&-& 6 \vD_{11\bar{1}} (\vg^a_{31}, \vu_c, \vu_c) 
- 6 \vDt_{11\bar{1}} (\vu_c, \vu_c, \vu_c) \rangle
\eea
whereas for the odd derivative term we obtain
\bea{s2-F}
\fl
\langle \vv_c| \vu_c \rangle d &:=& \langle \vv_c |
- \mL'(q_c) \vG_{31} +4\vC_{10} (\vu_c, \vg^a_{40}) 
+2\vC_{2\bar{1}} (\vG_{22}, \vg^a_{31}) \nonumber \\
\fl
&-& 4 \tilde{C}_{01} (\vG_{20}, \vu_c) 
- 2 \tilde{C}_{\bar{1}2} (\vu_c, \vG_{22}) 
+3 \vD_{11\bar{1}} (\vu_c, \vu_c, \vg^a_{31}) 
- 3 \tilde{D}_{\bar{1}11} (\vu_c, \vu_c, \vu_c) \rangle \quad .
\eea

All coefficients are real valued, since the constituents are real owing 
to the symmetry of the underlying system. The coefficients of the linear terms
can be expressed in terms of the spectrum. 
In addition, the cubic unfolding is obtained as a formal parameter derivative 
of the cubic coefficient of the ordinary Ginzburg Landau equation. One should
note that this contribution 
and the linear derivative term are both caused by the
parameter variation along the codimension one bifurcation manifold and are
easily missed if the parameters are not unfolded according to 
eq.(\ref{s2-h}). The mentioned properties are not passed
to the amplitude equation (\ref{s1-b}), since the coefficients are
renormalised by eqs.(\ref{s2-xa}).
For the remaining coefficients no simple interpretation seems to
be available.

\section{Properties of the amplitude equation}
Partial discussions of eq.(\ref{s1-b}) from different points of view can
be found in the literature \cite{DoEc}. Here we focus on those results 
which have in
our opinion consequences for the behaviour near the
codimension two bifurcation point.

First of all the coefficients $D$ and $s$ have to be positive respectively
negative in order to yield a bounded solution. We confine the subsequent
analysis to this case. Hence these coefficients can
be incorporated in the length scale as well as the magnitude of $\bar{A}$ 
and an additional parameter can be eliminated by a rescaling of the time.
However, since no real simplification is achieved we discuss the unscaled
equation directly.
\subsection{Potential case}
In the absence of the odd derivative term, $d=0$, eq.(\ref{s1-b}) admits
a potential $L=\int {\ell } dx$ decreasing in time,
with density
\bea{s3-a}
\fl
\ell &:=& -\bar{\eta} |\bar{A}|^2 + D |\partial_{\xi} \bar{A}|^2 
- \bar{r}/2 |\bar{A}|^4 
- s/3 |\bar{A}|^6
-\i g/4 |\bar{A}|^2 
\left( \bar{A}^* \partial_{\xi} \bar{A} 
- \bar{A} \partial_{\xi}\bar{A}^*\right)
\nonumber\\
\fl
&=& -\bar{\eta} |\bar{A}|^2 
+ D\left| \partial_{\xi} \bar{A} +\i g/(4D) |\bar{A}|^2 \bar{A} \right|^2  
- \bar{r}/2 |\bar{A}|^4 
-\left[s/3+g^2/(16 D)\right] |\bar{A}|^6 \quad .
\eea
The potential is definite for $s<-3 g^2/(16 D)$, so that in some sense
every solution tends to a time independent state. However,
if the inequality is violated, e.~g.~if $g$ is to large, the solutions may 
diverge. The potential property seems to be destroyed, if an
odd derivative term is present.
\subsection{Bifurcation scenario}
The trivial state $\bar{A}=0$ of the amplitude equation is stable if 
$\bar{\eta}<0$. Beyond this threshold time independent
plane waves emerge from this solution. For the existence of these solutions
$\bar{A}=\alpha_{\kappa} \exp(\i {\kappa} \xi)$ we obtain from eq.(\ref{s1-b})
the condition
\be{s3-b}
0=\bar{\eta}-D\kappa^2+(\bar{r}-\Delta \kappa) |\alpha_{\kappa}|^2
+ s |\alpha_{\kappa}|^4 \quad .
\ee
The quantity $\Delta:=g-d$ completely incorporates the dependence 
on the normal and the odd derivative term. 

\begin{figure}
\begin{center}
\mbox{\epsffile{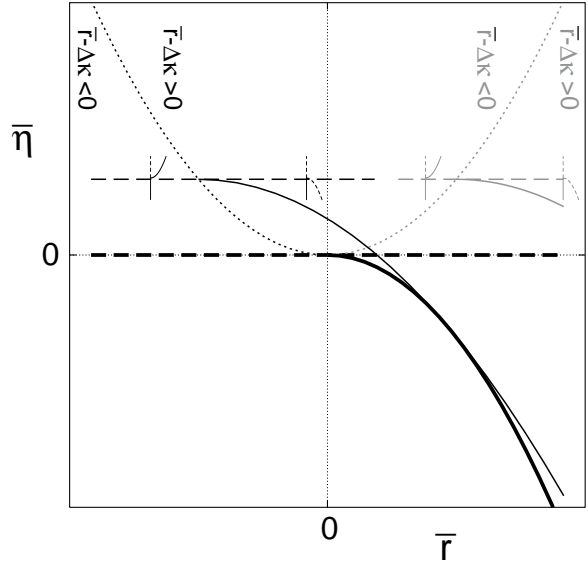}}
\end{center}
\caption[ ]{Sketch of a partial bifurcation diagram of the amplitude
equation (\ref{s1-b}) in the $\bar{r}$--$\bar{\eta}$ plane.
The thick broken line indicates the instability of the trivial state.
The thin broken line represents the bifurcation of
plane waves from the trivial state for a wave number
$\kappa$ with $\Delta \kappa >0$ (cf.~eq.(\ref{s3-c})).
The pitchfork like inserts indicate
whether the bifurcations are sub-- or supercritical and the thin dotted line
gives the transition for all wavenumbers (cf.~eq.(\ref{s3-e})). 
The thin solid line marks the saddle node bifurcation 
for the plane wave (cf.~eq.(\ref{s3-g})). 
The corresponding features for the wavenumber $-\kappa$ are displayed 
in grey. 
Finally the thick solid line marks the envelope of saddle node
bifurcation lines for all wavenumbers 
(cf.~eq.(\ref{s3-i})). \label{fig-b}}
\end{figure}

Consider for the moment an arbitrary but fixed
wavenumber $\kappa$. Eq.(\ref{s3-b}) determines the bifurcations of
the corresponding plane wave. It is evident that at 
\be{s3-c}
\bar{\eta}=D \kappa^2
\ee
a wave emerges from the trivial solution, and it is generated in 
parameter space on that side of the bifurcation set where the inequality
\be{s3-d}
(\bar{\eta}-D \kappa^2)(\bar{r}-\Delta \kappa)<0
\ee
is valid. Hence this peculiar bifurcation changes from sub-- to supercritical
behaviour at $\bar{r}=\Delta \kappa$, i.~e.~at
\be{s3-e}
\bar{\eta}=D \bar{r}^2/\Delta^2 \quad .
\ee
Now we are considering larger amplitudes $\alpha_k$ and concentrate on the
case, where the waves exhibit a saddle node bifurcation. If we rewrite
eq.(\ref{s3-b}) in the form
\be{s3-f}
0=s \left[ \left(|\alpha_{\kappa}|^2-
\frac{\bar{r}-\Delta \kappa}{-2s}
\right)^2 -\left(\frac{\bar{r}-\Delta \kappa}{-2 s}\right)^2
-\frac{\bar{\eta}-D\kappa^2}{-s} \right]
\ee
we immediately recognise that a saddle node bifurcation occurs at
\be{s3-g}
\bar{\eta}=D \kappa^2 - (-s)\left(
\frac{\bar{r}-\Delta \kappa}{-2 s}\right)^2 \quad ,
\ee
provided that the inequality
\be{s3-h}
(\bar{r}-\Delta \kappa)/(-2 s)>0
\ee
holds. The content of eqs.(\ref{s3-c}), (\ref{s3-d}), (\ref{s3-e}),
(\ref{s3-g}), and (\ref{s3-h}) is summarised in fig.\ref{fig-b} for
a particular wavenumber $\kappa$. 
The region of existence of a plane wave solution, 
which is bounded by eq.(\ref{s3-g}) extends beyond the 
stability region of the trivial state.

We now have to perform the analysis presented above for every wavenumber
$\kappa$. The region where plane wave solutions exist is given by the
union of the regions described above. Hence its boundary is
determined by the envelope of the curves (\ref{s3-g}) for all wavenumbers.
It is easily computed as (cf.~fig.\ref{fig-b})
\be{s3-i}
\bar{\eta}= - D \bar{r}^2/[ 4 D (-s) -\Delta^2], \quad (\bar{r}>0),
\qquad \mbox{if }|\Delta|<\Delta_c:=2 \sqrt{D (-s)} \quad .
\ee
If $|\Delta|$ approaches $\Delta_c$ the parabola (\ref{s3-i}) 
degenerates with the
negative $\bar{\eta}$--axis. For $|\Delta|>\Delta_c$ no boundary 
exists at all, so that for every $\bar{\eta}<0$ there exist plane 
wave solutions.

In summary, the scenario for $|\Delta|<\Delta_c$ 
resembles the sub--supercritical
transition in low dimensional dynamical systems, where a saddle--node 
bifurcation line is typically born. But in the extended case this behaviour
is destroyed at $|\Delta|=\Delta_c$, which can be viewed as a 
codimension three bifurcation point.

Yet we have not claimed the stability of the plane wave
solutions. A linear stability analysis according to
$\bar{A}=\alpha_k \exp(\i \kappa \xi) (1 + \db)$ yields
the linear equation
\bea{s3-j}
\fl 
\partial_{\tau} \db &=&
\bar{\eta} \db + D( - \kappa^2 \db + \i 2 \kappa \partial_{\xi} \db
+\partial_{\xi}^2 \db) + \bar{r}|\alpha_{\kappa}|^2 (\db^* + 2 \db)
+ s |\alpha_{\kappa}|^4 ( 2 \db^* + 3 \db) \nonumber \\
\fl
&-& (g-d) \kappa |\alpha_{\kappa}|^2 (2 \db + \db^*)
+ \i g |\alpha_{\kappa}|^2 \partial_{\xi} \db
+ \i d |\alpha_{\kappa}|^2 \partial_{\xi} \db^* \quad .
\eea
{  Splitting into real and imaginary parts and 
and taking the fixed point equation (\ref{s3-b}) into account
the corresponding real two dimensional system reads
\be{s3-ja}
\fl
\partial_{\tau}
\left( \begin{array}{c} \Re\delta \beta \\ \Im\delta \beta \end{array}\right)
=\left(\begin{array}{cc} 2(\bar{r}-\Delta \kappa) |\alpha_{\kappa}|^2
+ 4 s |\alpha_{\kappa}|^4 + D\partial_{\xi}^2 & (-2D\kappa -\Delta 
|\alpha_{\kappa}|^2)\partial_{\xi} \\
(2 D\kappa+(g+d)|\alpha_{\kappa}|^2)\partial_{\xi} & D \partial_{\xi}^2
\end{array} \right)
\left( \begin{array}{c} \Re\delta \beta \\ \Im\delta \beta \end{array}\right)
\quad .
\ee
Analysing the stability in terms of plane waves
$\exp(\i \omega \xi)$ yields}
a two--dimensional eigenvalue problem, where the trace and
the determinant of the corresponding matrix read
\bea{s3-k}
\Tr_{\omega} &=& 2\left[-D \omega^2 + (\bar{r}-\Delta \kappa)
|\alpha_{\kappa}|^2 + 2 s |\alpha_{\kappa}|^4\right] \label{s3-ka}\\
\Det_{\omega} &=& \omega^2\left[ D^2 \omega^2-
2D\left(\bar{r}-\Delta \kappa\right)|\alpha_{\kappa}|^2 - 4 D s
|\alpha_{\kappa}|^4 \right. \nonumber\\
&-& \left.\left(2 D \kappa+(g+d)|\alpha_{\kappa}|^2\right)
\left(2D \kappa + \Delta |\alpha_{\kappa}|^2\right)\right] \quad .
\eea
Stability
requires $\Tr_{\omega}<0$ and $\Det_{\omega}\ge 0$ for all wavenumbers
$\omega$. Owing to the simple dependence on $\omega$ the condition on
the trace results in
\be{s3-l}
|\alpha_{\kappa}|^2 > (\bar{r}-\Delta \kappa)/(-2s) \quad ,
\ee
which is of course valid if the right hand side is negative. Whenever the
right hand side is positive and the parameters are such that plane wave
solutions are possible, i.~e.~we are beyond the saddle node bifurcation line
(cf.~eqs.(\ref{s3-g}), (\ref{s3-h})), then eq.(\ref{s3-f}) tells us, that the
solution with the larger amplitude obeys the constraint (\ref{s3-l}), whereas
the solution with the smaller amplitude is unstable. Hence we are left with 
checking the condition on the determinant which results in
\be{s3-m}
\fl
4 D (-s) |\alpha_{\kappa}|^2 \left[ |\alpha_{\kappa}|^2
- (\bar{r}-\Delta \kappa)/(-2s)\right] 
-\left(2 D \kappa+(g+d)|\alpha_{\kappa}|^2\right)
\left(2D \kappa + \Delta |\alpha_{\kappa}|^2\right) \ge 0 \quad .
\ee
Due to this condition it is evident, that the stability properties depend
on the parameters $g$ and $d$ separately. 
A complete discussion of the stability using eq.(\ref{s3-m})
is straightforward but tedious. We do not
intend to discuss the full implications of this inequality, but just 
concentrate on a neighbourhood of the envelope (\ref{s3-i}) { 
in order to study whether stable solutions are generated at this 
bifurcation line}. 
For that purpose we fix the wavenumber to 
$\kappa=-\Delta \bar{r}/(\Delta_c^2-\Delta^2)$ which is just the value
for which the saddle node bifurcation line touches the envelope
(cf.~fig.\ref{fig-b} and eqs.(\ref{s3-f}), (\ref{s3-g}), and (\ref{s3-i}))
and expand the left hand side of eq.(\ref{s3-m}) for $\bar{\eta}$ values
slightly beyond the envelope (cf.~eq.\ref{s3-g}). We obtain to the
leading order the result
\be{s3-ma}
\left[4 D (-s) - 2 g \Delta\right ] |\alpha_{\kappa}|^2 \left[
|\alpha_{\kappa}|^2 - (\bar{r}-\Delta \kappa)/(-2 s)\right] \ge 0 \quad .
\ee
For the solution with the larger amplitude (cf.~eq.(\ref{s3-l})) the
condition is satisfied provided that $g \Delta < 2 D (-s)$ holds. 
Then a stable plane wave occurs at the envelope.

\section{Conclusion}
We have presented the complete and systematic derivation of the 
amplitude equation, which 
governs the transition from sub-- to supercritical soft--mode instabilities.
Within our approach the general expression for the coefficients 
and especially their dependence on the system parameters has been obtained.
Although these formulas look a little bit lengthy, one should keep in mind
that they can be applied to almost every physical situation, and
that their evaluation is straightforward in concrete cases. 

From the principal point of view it is worth to mention, that the amplitude
equation (\ref{s1-b}) can be derived consistently at all. Such a feature
is far from being obvious. To emphasise this point consider the corresponding 
hard--mode case, i.~e.~an instability at $q_c=0$ with a nonvanishing
frequency. A superficial inspection would suggest that the whole 
derivation goes along the same lines with minor modifications. But if we
follow the approach of section \ref{sec2}, we are left at the third order
with eq.(\ref{s2-q}). Since now all expressions are complex valued but
the higher order codimension condition requires a vanishing real part only,
the secular condition becomes a nonlinear equation. Of course it can be
easily integrated to yield the time dependence on the scale 
$\tau_1$ as $A=\tilde{A} \exp[\i \Im (\rho_3) |\tilde{A}|^2 \tau_1]$. Here
the constant of integration $\tilde{A}$ depends on the slower scales. 
If one uses this representation in
the subsequent orders, then the derivatives with respect to spatial
coordinates yield linearly in $\tau_1$ increasing terms, since the exponent
is space dependent. This feature invalidates the systematic derivation
although a semi--quantitative approach has been proposed (cf.~the
discussion in \cite{BLN}).

The origin of such difficulties lies in high frequency components which 
contribute to the secular conditions in low orders and cause an 
uncontrolled mixing of different time scales.
Similar phenomena are well known in the problem of
counterpropagating waves, 
where a formal derivation is still possible and
a nonlocal coupling in the amplitude equation is generated \cite{KnLu}.
For the degenerated hard--mode instability we expect finally
similar effects
but further investigations are needed. Nevertheless these considerations
emphasise again
the necessity of careful derivations of amplitude equations
to supplement phenomenological approaches.

Concerning the behaviour beyond the sub--supercritical soft--mode
instability we stress that without an odd derivative term a
potential system occurs. Hence that kind of term is responsible
for a persistent time evolution beyond the threshold. In addition, we
mention that the difference $\Delta$ between the normal and the odd 
derivative term determines the domain of existence of spatially periodic
patterns in the vicinity of the threshold. These properties again show
that the behaviour beyond the instability depends crucially
on the actual numerical values of the coefficients.
\appendix
\section{Inhomogeneous part}\label{appa}
For convenience we list in this appendix the inhomogeneous parts, which
occur in each stage of the derivation of the amplitude equation.
For notational simplicity the same label is assigned to
the components at each order and we use an overbar to denote negative indices
$\bar{n}=-n$. All the abbreviations which we introduce are real valued.

{  If we insert eq.(\ref{s2-i}) into the equation of motion 
(\ref{s2-a}) and observe the expansion (\ref{s2-j}) we obtain at the
order $\Or(\eps^{1/2})$
\be{a-aa}
0=\mL(q_c) \vu_c \exp(i q_c x) A + c.c.
\ee
since the spatial derivatives act on the plane wave only. This condition
is fulfilled by virtue of the eigenvalue equation for the critical mode
(\ref{s2-ca}).

At the order $\Or(\eps)$ one obtains a contribution from the temporal and
spatial derivatives acting on $\vPhi_2$ and one contribution from the
quadratic nonlinearity (\ref{s2-ea}) at $\vmu=\vmu_c$
with the derivatives acting only on the
plane waves. Taking the abbreviation (\ref{s2-ga}) into account the result
reads
\bea{a-ab}
\partial_t \vPhi_2 &=& {\cal L} \vPhi_2 
+\sum_{\alpha \beta} C^{(\alpha \beta)} \{ \partial_x^{\alpha} \vPhi_1,
\partial_x^{\beta} \vPhi_1 \} \nonumber\\
&=&
{\cal L} \vPhi_2 + \vC_{1\bar{1}}(\vu_c,\vu_c)
|A|^2 +\left[\vC_{11}(\vu_c,\vu_c) \exp(2 i q_c x) A^2 + c.c.\right] \quad .
\eea
Hence in the notation of eq.(\ref{s2-l})}
the nonvanishing Fourier components read
\be{a-a}
\vw_0= -2 \mL(0) \vG_{2 0} |A|^2, \quad \vw_2= - \mL(2 q_c) \vG_{2 2} A^2
\quad ,
\ee
where the abbreviations
\be{a-b}
\vG_{20}:=-\mL(0)^{-1} \vC_{1\bar{1}}(\vu_c,\vu_c),\quad
\vG_{22}:=-\mL(2 q_c)^{-1} \vC_{11}(\vu_c,\vu_c)
\ee
have been used, and the obvious relation $\vw_{\bar{n}}=\vw_n^*$ should 
be noted. {  The solution of the linear equation (\ref{a-ab}), 
discarding exponentially
decaying transients, is given by eq.(\ref{s2-p}).

At the order $\Or(\eps^{3/2})$ the time derivative and the linear
operator act plainly on $\vPhi_3$. In addition, these derivatives
give a contribution when acting on the slow scales of $\vPhi_1$.
For the nonlinearities now the quadratic and the cubic 
terms at $\vmu=\vmu_c$ contribute 
\bea{a-ba}
\partial_t \vPhi_3 &=& {\cal L} \vPhi_3 + {\cal L}^{(1)}\vPhi_1+
({\cal L} \vPhi_1)^{[1]}- \partial_{\tau_1} \vPhi_1\nonumber\\
&+& \sum_{\alpha \beta} 2 C^{(\alpha \beta)}
\{ \partial_x^{\alpha} \vPhi_2,\partial_x^{\beta} \vPhi_1\}
+ \sum_{\alpha \beta \gamma} D^{(\alpha \beta \gamma)}
\{\partial_x^{\alpha} \vPhi_1,\partial_x^{\beta} \vPhi_1,
\partial_x^{\gamma} \vPhi_1\} \quad .
\eea
Here the notation $({\cal L} \vPhi_1)^{[1]}$ means that the derivatives 
have to be evaluated at
the first order in $\eps$. All other derivatives with respect to $x$
are understood at fixed values of $\xi_k$. Taking the relation
$(\partial_x^{\alpha} \exp(i q_c x) A)^{[1]}= \alpha (i q_c)^{\alpha-1}
\exp(i q_c x) \partial_{\xi_1}A$ into account the third contribution 
on the right hand
side is expressed in terms of the derivative of the matrix (\ref{s2-c})
with respect to $k$. If we evaluate the
nonlinear contribution with the help of the solution (\ref{s2-p}) of
the preceding order and recast
all contributions into the form (\ref{s2-l}) we get}
\bea{a-c}
\vw_0 &=&-2  \mL(0) \vG_{2 0} A B^* \nonumber\\ 
\vw_1 &=&
-\vu_c \partial_{\tau_1} A +  
\mL^{(1)}(q_c)\vu_c A- 
\i  \mL'(q_c) \vu_c \partial_{\xi_1} A \nonumber\\
&+& \left[ 2 \vC_{2\bar{1}} (\vG_{22}, \vu_c) +
4\vC_{10} (\vu_c, \vG_{20})+ 3\vD_{11\bar{1}} (\vu_c, \vu_c, \vu_c)  
\right] |A|^2 A \nonumber \\
\vw_2&=&- 2 \mL(2 q_c) \vG_{2 2 } A B \nonumber\\
\vw_3&=& - \mL(3 q_c) \vG_{3 3} A^3
\eea
with the abbreviation
\be{a-d}
\vG_{3 3} := -\mL^{-1}(3 q_c) \left[
2\vC_{21} (\vG_{22}, \vu_c) + \vD_{111} (\vu_c, \vu_c, \vu_c)\right]
\ee
Here $\mL'(k)$ denotes the derivative with respect to $k$.
{  The nonsecular solution discarding transients is given by 
eq.(\ref{s2-t}) where the abbreviations 
\bea{a-de}
\vG_{3 1} := -\mL_c^{-1} \left[ 2\vC_{2\bar{1}} (\vG_{22}, \vu_c) +
4\vC_{10} (\vu_c, \vG_{20}) + 3\vD_{11\bar{1}} (\vu_c, \vu_c, \vu_c)  
\right]
\label{a-da}\\
\vg_{31}^a := -\mL_c^{-1} \mL'(q_c) \vu_c = 
(\Id -|\vu_c\rangle \langle \vv_c|)
\partial_k \vu_k (\vmu_c)|_{k=q_c}
\label{a-dc}\\
\vg_{31}^b := -\mL_c^{-1} \mL^{(1)}(q_c) \vu_c = 
(\Id -|\vu_c\rangle \langle \vv_c|)
(\vmu^{(1)} \partial_{\vmu}) \vu_{q_c} (\vmu)|_{\vmu=\vmu_c} \quad .
\label{a-dd}
\eea
have been introduced.

Proceeding to the order $\Or(\eps^2)$ one has to observe that in
addition the parameter dependence of the quadratic term has to be
taken into account (cf.~eq.(\ref{s2-h})). Denoting
the corresponding directional derivative by
$C^{(\alpha\beta)(1)}=(\vmu^{(1)}\partial_{\vmu})
C^{(\alpha\beta)}_{\vmu}|_{\vmu=\vmu_c}$ the equation reads
\bea{a-df}
\fl
\partial_t \vPhi_4 &=& {\cal L} \vPhi_4 + {\cal L}^{(1)}\vPhi_2+
({\cal L} \vPhi_2)^{[1]}- \partial_{\tau_1} \vPhi_2\nonumber\\
\fl
&+& \sum_{\alpha \beta} \left[
C^{(\alpha \beta)}
\{ \partial_x^{\alpha} \vPhi_2,\partial_x^{\beta} \vPhi_2\}+
2 C^{(\alpha \beta)}
\{ \partial_x^{\alpha} \vPhi_3,\partial_x^{\beta} \vPhi_1\}\right. \nonumber\\ 
\fl & & \left. +\,
2 C^{(\alpha \beta)}
\{ (\partial_x^{\alpha} \vPhi_1)^{[1]},\partial_x^{\beta} \vPhi_1\}+
C^{(\alpha \beta)(1)}
\{ \partial_x^{\alpha} \vPhi_1,\partial_x^{\beta} \vPhi_1\}\right] \nonumber\\
\fl
&+& \sum_{\alpha \beta \gamma} 3 D^{(\alpha \beta \gamma)}
\{\partial_x^{\alpha} \vPhi_2,\partial_x^{\beta} \vPhi_1,
\partial_x^{\gamma} \vPhi_1\}
+ \sum_{\alpha \beta \gamma \delta} E^{(\alpha \beta \gamma \delta)}
\{ \partial_x^{\alpha} \vPhi_1,\partial_x^{\beta} \vPhi_1,
\partial_x^{\gamma} \vPhi_1,\partial_x^{\delta} \vPhi_1 \} \quad .
\eea
Inserting the solutions of the preceding orders and performing the
derivatives we obtain for} the Fourier modes
up to wavenumber $2 q_c$
\bea{a-e}
\fl
\vw_0&=& \mL(0) \left[ -2 \vG_{20} \left(
A C^* + A^* C + |B|^2\right) - 2 \i \vg_{40}^a \left(
A \partial_{\xi_1} A^* - A^* \partial_{\xi_1} A \right) \right.
\nonumber\\
\fl
&-& \left. \vG_{40} |A|^4
- 2 \vg_{40}^b |A|^2 \right] \nonumber\\
\fl
\vw_1&=& - \vu_c \partial_{\tau_1} B 
+ \mL^{(1)}(q_c) \vu_c B
- \i \mL'(q_c) \vu_c \partial_{\xi_1} B \nonumber\\
\fl &+&   
\left[ 2\vC_{2\bar{1}} (\vG_{22}, \vu_c) + 4\vC_{10} (\vu_c, \vG_{20}) +
3\vD_{11\bar{1}} (\vu_c, \vu_c, \vu_c) \right] 
\left[ 2 |A|^2 B + A^2 B^* \right] \nonumber \\
\fl
\vw_2&=&
-\mL(2 q_c)\left[ \vg_{42}^b A^2 + 2 \vG_{22} A C  
+2 \i \vg^a_{42} A \partial_{\xi_1} A +\vG_{42} A^* A^3 +\vG_{22} B^2 \right]
\eea
with the abbreviations
\bea{a-f}
\fl
\vG_{40} &:=& -\mL(0)^{-1}
\left[2\vC_{1\bar{1}} (\vG_{31}, \vu_c) +
2\vC_{1\bar{1}} (\vu_c, \vG_{31}) +
2\vC_{2\bar{2}} (\vG_{22}, \vG_{22}) 
+4\vC_{00} (\vG_{20}, \vG_{20}) \right. \nonumber \\
\fl &+ &
3\vD_{11\bar{2}} (\vu_c, \vu_c, \vG_{22}) +
3\vD_{2\bar{1}\bar{1}} (\vG_{22}, \vu_c, \vu_c)
+12\vD_{10\bar{1}} (\vu_c, \vG_{20}, \vu_c) \nonumber\\
\fl &+& \left. 
6 \vE_{11\bar{1}\bar{1}}(\vu_c,\vu_c,\vu_c,\vu_c)\right] \label{a-fa}\\
\fl
\vg^a_{40} &:=&  -\mL(0)^{-1}\ \left[ 
-\mL'(0) \vG_{20} +  \vC_{1\bar{1}} (\vu_c, \vg^a_{31}) -
\vCt _{\bar{1}1} (\vu_c, \vu_c)\right] \label{a-fb}\\ 
\fl
\vg^b_{40} &:=& -\mL(0)^{-1} 
\left[ \vC_{1\bar{1}} (\vg^b_{31}, \vu_c) +
\vC_{1\bar{1}} (\vu_c, \vg^b_{31}) +  \mL^{(1)}(0)\ \vG_{20} 
+ \vC_{1\bar{1}}^{(1)} (\vu_c, \vu_c)  \right] \label{a-fc}\\
\fl
\vG_{42} &:=& -\mL(2 q_c)^{-1}
\left[ 2 \vC_{11} (\vG_{31}, \vu_c) + 
2 \vC_{3\bar{1}} (\vG_{33}, \vu_c) + 4 \vC_{20} (\vG_{22}, \vG_{20})
\right. \nonumber\\
\fl &+& \left.
6\vD_{110} (\vu_c, \vu_c, \vG_{20}) +
6 \vD_{21\bar{1}} (\vG_{22}, \vu_c, \vu_c) +
6 \vE_{111\bar{1}} (\vu_c,\vu_c,\vu_c,\vu_c) \right] \label{a-fd} \\
\fl
\vg^a_{42} &:=&  -\mL(2 q_c)^{-1} \left[
- \mL'(2 q_c) \vG_{22} -  \vC_{11} (\vg^a_{31}, \vu_c)  
- \vCt_{11} (\vu_c, \vu_c) \right] \label{a-fe} \\ 
\fl
\vg^b_{42} &:=&  - \mL(2 q_c)^{-1} \left[ 
2 \vC_{11} (\vu_c, \vg^b_{31}) + \mL^{(1)} (2 q_c) \vG_{22} + 
\vC_{11}^{(1)} (\vu_c, \vu_c) \right] \label{a-ff}
\eea
and
\be{a-g}
\vCt_{m n}(\vu, \vv) := \i \sum_{\alpha \beta} \alpha (\i m q_c)^{\alpha -1}
(\i n q_c)^{\beta} \vC^{(\alpha \beta)}\{ \vu, \vv\} \quad .
\ee
{   The nonsecular solution of eq.(\ref{a-df}) up to Fourier 
modes $\pm 2q_c$ is then given by eq.(\ref{s2-w}).}
We remark that, if the definitions (\ref{a-b}) are understood in terms of 
the full parameter dependent quantities and eigenvectors  
(cf.~eqs.(\ref{s2-c}), (\ref{s2-ga}), and (\ref{s2-ca})), 
then the abbreviations (\ref{a-fc}) and (\ref{a-ff})
obey, taking relation (\ref{a-dd}) into account
\bea{a-h}
\vg_{40}^b &=& \left. (\vmu^{(1)} \partial_{\vmu}) 
\vG_{20}\right|_{\vmu=\vmu_c}
-\vG_{20} \left(\langle \vv_c | (\vmu^{(1)}\partial_{\vmu}) 
\vu _{q_c}(\vmu)|_{\vmu=\vmu_c}\rangle + c.c.\right)\label{a-ha}\\
\vg_{42}^b &=& \left.
(\vmu^{(1)} \partial_{\vmu}) \vG_{22}\right|_{\vmu=\vmu_c}
-2 \vG_{22} \langle \vv_c | (\vmu^{(1)}\partial_{\vmu})
\vu_{q_c}(\vmu)|_{\vmu=\vmu_c}\rangle \label{a-hb} \quad .
\eea

{  Finally at the order $\Or(\eps^{5/2})$ we end up with
\bea{a-i}
\fl
\partial_t \vPhi_5 &=& {\cal L} \vPhi_5 + {\cal L}^{(1)}\vPhi_3+
{\cal L}^{(2)}\vPhi_1+
({\cal L} \vPhi_3)^{[1]}- \partial_{\tau_1} \vPhi_3+
({\cal L} \vPhi_1)^{[2]}- \partial_{\tau_2} \vPhi_1+ 
({\cal L}^{(1)} \vPhi_1)^{[1]}
\nonumber\\
\fl
&+& \sum_{\alpha \beta} \left[
2 C^{(\alpha \beta)}
\{ \partial_x^{\alpha} \vPhi_4,\partial_x^{\beta} \vPhi_1\}+
2 C^{(\alpha \beta)}
\{ \partial_x^{\alpha} \vPhi_3,\partial_x^{\beta} \vPhi_2\}\right. \nonumber\\ 
\fl & & \left. +\,
2 C^{(\alpha \beta)}
\{ (\partial_x^{\alpha} \vPhi_2)^{[1]},\partial_x^{\beta} \vPhi_1\}+
2 C^{(\alpha \beta)}
\{ (\partial_x^{\alpha} \vPhi_1)^{[1]},\partial_x^{\beta} \vPhi_2\}+
2 C^{(\alpha \beta)(1)}
\{ \partial_x^{\alpha} \vPhi_2,\partial_x^{\beta} \vPhi_1\}\right] \nonumber\\
\fl
&+& 
\sum_{\alpha \beta \gamma} \left[ 
3 D^{(\alpha \beta \gamma)}
\{\partial_x^{\alpha} \vPhi_3,\partial_x^{\beta} \vPhi_1,
\partial_x^{\gamma} \vPhi_1\}+
3 D^{(\alpha \beta \gamma)}
\{\partial_x^{\alpha} \vPhi_2,\partial_x^{\beta} \vPhi_2,
\partial_x^{\gamma} \vPhi_1\} \right. \nonumber\\
\fl & & \left. +\,
3 D^{(\alpha \beta \gamma)}
\{(\partial_x^{\alpha} \vPhi_1)^{[1]},\partial_x^{\beta} \vPhi_1,
\partial_x^{\gamma} \vPhi_1\}+
D^{(\alpha \beta \gamma)(1)}
\{\partial_x^{\alpha} \vPhi_1,\partial_x^{\beta} \vPhi_1,
\partial_x^{\gamma} \vPhi_1\}\right] \nonumber\\
\fl
&+& \sum_{\alpha \beta \gamma \delta} 4 E^{(\alpha \beta \gamma \delta)}
\{ \partial_x^{\alpha} \vPhi_2,\partial_x^{\beta} \vPhi_1,
\partial_x^{\gamma} \vPhi_1,\partial_x^{\delta} \vPhi_1 \}\nonumber\\
\fl &+& \sum_{\alpha \beta \gamma \delta \epsilon} 
F^{(\alpha \beta \gamma \delta \epsilon)}
\{ \partial_x^{\alpha} \vPhi_1,\partial_x^{\beta} \vPhi_1,
\partial_x^{\gamma} \vPhi_1,\partial_x^{\delta} \vPhi_1,
\partial_x^{\epsilon} \vPhi_1 \} \quad .
\eea
As before $D^{(\alpha \beta \gamma)(1)}$ denotes the directional derivative 
and $({\cal L}\vPhi_1)^{[2]}$ indicates
that spatial derivatives have to be evaluated at order $\Or(\eps^2)$.
Inserting the previous orders and performing the derivatives
the Fourier mode $\vw_1$ of
the inhomogeneous part of eq.(\ref{a-i}) is evaluated and yields 
after some algebra the secular condition (\ref{s2-x}), taking the
abbreviation
\be{a-j}
\vDt_{l m n}(\vu,\vv,\vw):= \i \sum_{\alpha \beta \gamma}
\alpha (\i l q_c)^{\alpha -1} (\i m q_c)^{\beta} (\i n q_c)^{\gamma}
\vD^{(\alpha \beta \gamma)}\{\vu,\vv,\vw\}
\ee
into account.}


\section*{References}
\begin{thebibliography}{99}
\bibitem{CrHo} Cross M C and Hohenberg P C 1993 \RMP {\bf 65} 851
\bibitem{GuHo} Guckenheimer J and Holmes P 1986 {\it Nonlinear Oscillations,
Dynamical Systems, and Bifurcations of Vector Fields} (New York: Springer)
\bibitem{Mann} Manneville P 1990 {\it Dissipative Structure and Weak
Turbulence} (San Diego: Acad.~Press)
\bibitem{Coll} Collet P 1994 \NL {\bf 7} 1175 
\bibitem{BeSw} Becerril R and Swift J B 1997 \PR E {\bf 55} 6270
\bibitem{DeBr} Deissler R J and Brand H R \PL A 1990 {\bf 146} 252
\bibitem{SaHo} van Saarloos W and Hohenberg P C 1992 {\it Physica} D {\bf 56}
303; Erratum 1993 {\it Physica} D {\bf 69} 209
\bibitem{EcIo} Eckhaus W and Iooss G 1989 {\it Physica} D {\bf 39} 124
\bibitem{DoEc} Doelman A and Eckhaus W 1991 {\it Physica} D {\bf 53} 249
\bibitem{BLN} Brand H, Lomdahl P S, and Newell A C \PL A 1986 {\bf 118} 67
\bibitem{KnLu} Knobloch E and de Luca J 1990 \NL {\bf 3} 975
\end{thebibliography}
\end{document}